\documentclass[preprint,12pt,authoryear]{elsarticle}
\usepackage{fullpage}
\usepackage{array}
\usepackage[caption]{subfig} 
\usepackage{hyperref,enumitem}
\usepackage{amsmath}
\usepackage{rotating}
\usepackage{amsfonts}
\usepackage{multicol} 
\usepackage{framed}
\usepackage{algorithm,algorithmic}
\usepackage{natbib}
\usepackage{graphicx}
\usepackage{rotating}
\usepackage{booktabs,framed,placeins}
\usepackage{epsfig}
\usepackage{float}
\usepackage{xcolor}
\usepackage{slashbox}
\usepackage{adjustbox}

\newtheorem{thm}{Theorem}[subsection]
\newtheorem{definition}[thm]{Definition}

\usepackage{mathtools}

\newcommand{\javaplex}{\textsc{javaPlex}}
\newcommand{\perseus}{\textsc{Perseus}}
\newcommand{\gudhi}{\textsc{Gudhi}}

\newcommand{\dionysus}{\textsc{Dionysus}}

\newcommand{\ripser}{\textsc{ripser}}

\DeclareMathOperator{\image}{Image}
\DeclareMathOperator{\kernel}{Kernel}

\usepackage[normalem]{ulem}

\usepackage{amssymb}

\begin{document}

\begin{frontmatter}


\title{Classification of COVID-19 via Homology of CT-SCAN}

\author[ad1]{Sohail Iqbal\corref{cor1}\fnref{fn1} }
\ead{soh.iqbal@gmail.com}
\cortext[cor1]{Corresponding author}

\author[ad1]{Hafiz Fareed Ahmed\fnref{fn1} }
\ead{fa.usmani@gmail.com}

\address[ad1]{Department of Mathematics, COMSATS University Islamabad, Park Road, Islamabad, Pakistan}


\author[ad2]{Talha Qaiser \fnref{fn1}}
\ead{t.qaiser@imperial.ac.uk }
\address[ad2]{Department of Computing, Imperial College London, SW72AZ, United Kingdom}

\fntext[fn1]{These authors contributed the same during this project}

\author[ad3]{Muhammad Imran Qureshi}
\ead{Imran.qureshi@kfupm.edu.sa}
\address[ad3]{Department of Mathematics, King Fahd University of Petroleum and Minerals (KFUPM), Dhahran, Saudi Arabia}

 \author[ad4]{Nasir Rajpoot}
 \ead{N.M.Rajpoot@warwick.ac.uk}
 \address[ad4]{Deparment of Computer Science, The University of Warwick, Coventry, CV47AL, United Kingdom}



\begin{abstract}
In this worldwide spread of SARS-CoV-2 (COVID-19) infection, it is of utmost importance to detect the disease at an early stage especially in the hot spots of this epidemic. There are more than 110 Million infected cases on the globe, sofar. Due to its promptness and effective results computed tomography (CT)-scan image is preferred to the reverse-transcription polymerase chain reaction (RT-PCR). Early detection and isolation of the patient is the only possible way of controlling the spread of the disease. Automated analysis of CT-Scans can provide enormous support in this process. In this article,
We propose a  novel approach to detect SARS-CoV-2 using CT-scan images. Our method is based on a very intuitive and natural idea of analyzing shapes, an attempt to mimic a professional medic. We mainly trace SARS-CoV-2 features by quantifying their topological properties.  We primarily use a tool called persistent homology, from Topological Data Analysis (TDA), to compute these topological properties.

We train and test our model on the ``SARS-CoV-2 CT-scan dataset" \citep{soares2020sars}, an open-source dataset,  containing 2,481 CT-scans of normal and COVID-19 patients. Our model yielded an overall benchmark F1 score of $99.42\% $, accuracy $99.416\%$, precision $99.41\%$, and recall $99.42\%$. The TDA techniques have great potential that can be utilized for efficient and prompt detection of  COVID-19.  The immense potential of TDA may be exploited in clinics for rapid and safe detection of COVID-19 globally, in particular in the low and middle-income countries where RT-PCR labs and/or kits are in a serious crisis.
\end{abstract}


\begin{graphicalabstract}
\begin{figure} [H]
        \centering
        \includegraphics[width=15cm]{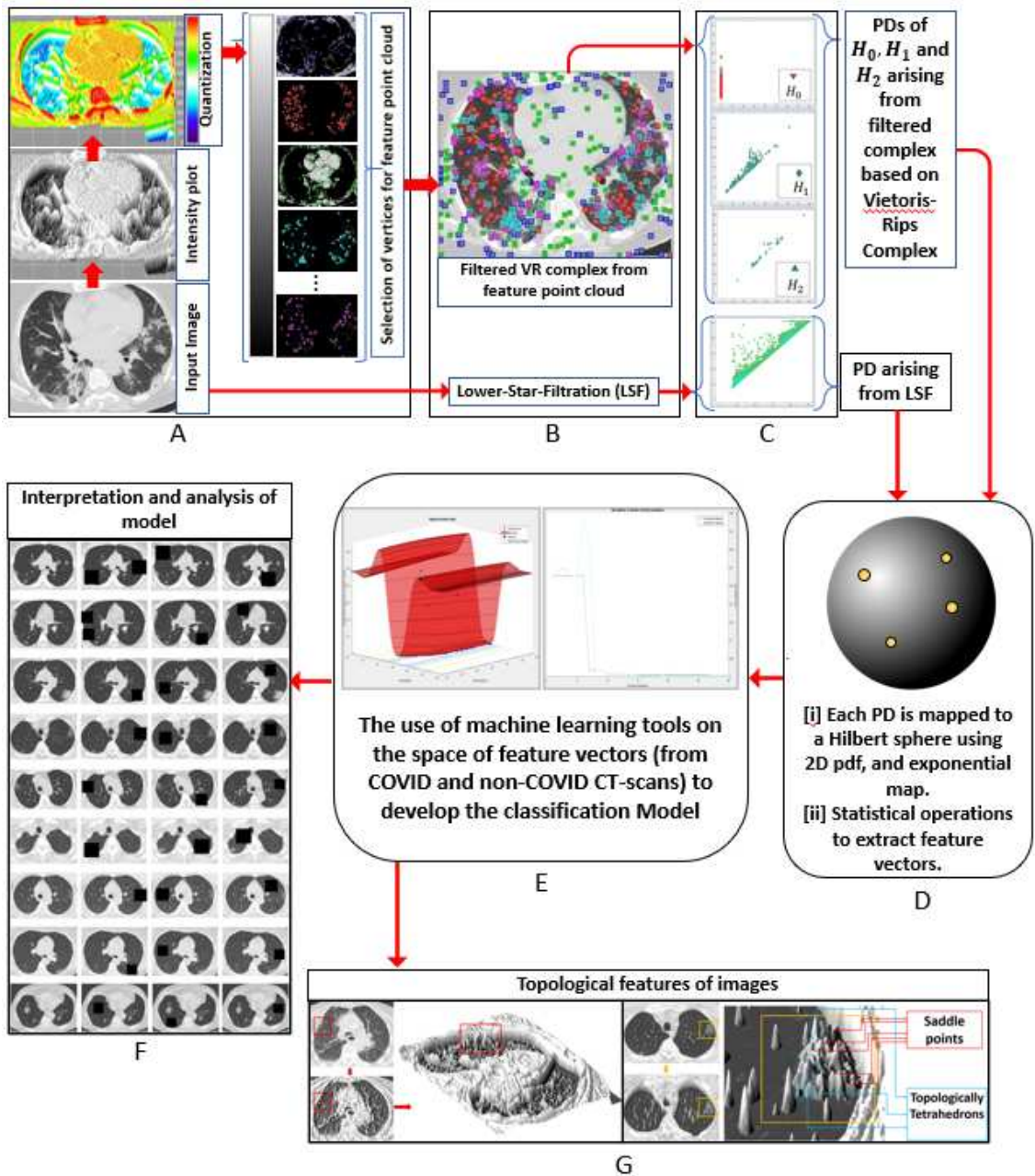}
        \caption{Graphical Abstract}
        \label{GraphicalAbstract}
\end{figure}
\end{graphicalabstract}





\end{frontmatter}


\section{Introduction}

On March 11 2020 WHO declared a pandemic caused by the novel corona virus (nCoV). The  disease is well known as SARS-CoV-2, or COVID-19, and originated from Wuhan, China in December 2019. After more than a year the virus is still spreading exponentially and has reached 212 countries. At the time of writing this paper there are more than 110 million infected cases, whereas the death toll has crossed 2.4 million.

The main reason for the spread of the virus is its asymptomatic nature, where an affected person  spreads the virus without any signs of illness. The only precaution that is required worldwide is  immediate isolation of the affected person. For this purpose, effective and timely testing is the core factor for treatment and  prevention. 

COVID-19 is a member of the family of viruses SARS (Severe Acute Respiratory Syndrome). It causes severe respiratory illness and primarily damages the lungs \citep{WuhanClinicalFeatures}. We use ``Reverse Transcription Polymerise Chain Reaction (RT-PCR)" and chest computed tomography (CT) scan  for the  detection of COVID-19. In RT-PCR  we reverse transcription of RNA into DNA, and then detect the presence of virus DNA. This method is a laboratory technique that requires a trained personnel that carries out the whole process. The test might take hours, if not days, to give the results. This situation worsens, for extremely affected areas with limited test kits and trained personnel. The RT-PCR testing also gives us the false-negative results, in some cases \citep{PCRAccuracy}, while in the chest CT-scan, the image analysis has been done by radiologists \citep{imaging}. The latter is much more sensitive  method than the RT-PCR. The study of 1014 cases shows the significance of chest CT-scans over RT-PCR \citep{correlation}.
One study finds that  40 out of 41 (98\%) patients had pneumonia with abnormal findings on chest CT-scans \citep{WuhanClinicalFeatures}.

The image analysis by radiologists suggest that in COVID-19-positive patients, the ground-glass opacities (GGOs) together with consolidations, crazy pavings appear at the peripheral portions of bilateral lungs. The increased attenuation in chest CT scan is the main feature in detection of COVID-19. Detection of these features is relatively   time efficient due to higher sensitivity of this method \citep{CTchanges}.

Over the past decade, one of the promising directions in health care innovation is the applicability of artificial intelligence (AI) in medical imaging. In recent years, AI in general has revolutionized the field of computer vision \citep{voulodimos2018deep}, and natural language processing \citep{ruder2019transfer,ruder2019neural} by pushing state-of-the-art performance in various pattern recognition tasks. More recently, there is an upward trend in exploring the usability of  machine learning algorithms for medical imaging data \citep{currie2019machine}. In the current scenario with the worldwide outbreak of SARS-CoV2, it is imperative to develop  screening tools to analyze the COVID-19 chest CT-scans. One of the main challenges in deep learning is data hungriness. In order to converge a deep learning model one may require thousands, and in some cases millions of images \citep{deng2009imagenet} for training. On the other hand techniques of Topological Data Analysis (TDA) use geometrical features, making them efficient in terms of amount of data required, speed,  predictability and interpretability.  

TDA is one of the rapidly growing techniques in data analysis. It provides tools to analyze data by bridging techniques from machine learning, statistics, algebraic topology,  topology, and algebra. One of the main tools in TDA is persistent homology (PH). It is a very effective technique that record the intrinsic topological properties of data. The essential idea is to produce topological features across a scale. On this scale, some features ``die'' early and some ``live'' longer. The persisting times of these features are the key point of PH. The ideas of  PH has been successfully applied to many areas of science and technology vis-\`{a}-vis network structures \citep{desilva2007, Lee2012}, computational biology \citep{ComptBiologybtm250, yao2009topological,wang2016object}, data analysis \citep{carlsson2009topology,liu2012fast,rieck2012multivariate}, image analysis \citep{carlsson2009computing,frosini2013persistent,bendich2010computing}, amorphous material structures \citep{hiraoka2016hierarchical}, etc.  

In the recent years these techniques has been successful in medical image analysis. For example, in \citep{QAISER20191,qaiser2016persistent,qaiser2017tumor} the authors developed models, based on PH, for efficient tumor segmentation in whole-slide images of histology slides; in \citep{garside2019topological} the topological features are used to differentiate  healthy patients and those with diabetic retinopathy; in \citep{chung2018topological} segmentation of skin cancer using a given image was achieved using these techniques, etc.

In this work we develop a state-of-the-art model to detect traces of COVID-19 infection in  CT-scans. To train and test our model we use ``SARS-CoV-2 CT-scan dataset" \citep{soares2020sars}. 

\begin{figure}[H] 
        \centering
        \fbox{ \includegraphics[width=15cm]{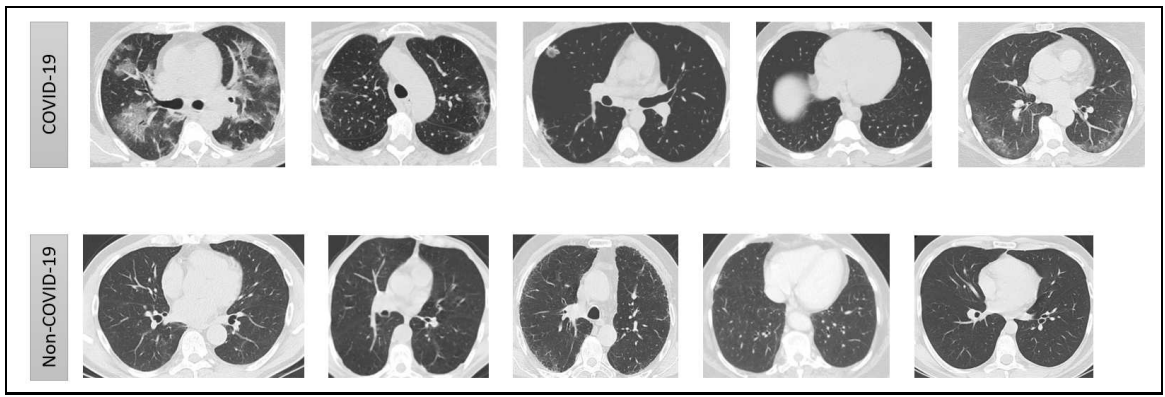}}
        \caption{ A sample of CT-Scan images from ``SARS-CoV-2 CT-scan dataset" for COVID-19 patients (first row) and normal cases (second row).}
        \label{samples}
\end{figure}
There are mainly three stages to develop our model. In the first stage, we devise a way to construct a simplicial complex from a give image and then calculate PD's associated to it. We map our PDs on a Hilbert sphere, following Anirudh et al \citep{anirudh2016riemannian}, in the second stage. This step enable us to perform different statistical operations on the space of PDs. In the last stage, we use SVM to develop our classification model. 

The rest of the paper is structured as follows. We review basics of PH in Section \ref{PH}. In Section \ref{MaA} we describe our methodology to extract features from CT-scans. Our restults are reported in \ref{RaD}. Finally, in Section \ref{Con}, we draw our conclusions. 

\section{Mathematical Preliminaries}\label{PH}
In this section we  recall basic notions and definitions leading to persistent homology and persistent diagrams (PD's).    

\subsection{Persistent homology (PH)}
The persistent homology is one of the main tools used in topological data analysis (TDA). It provides a way to analyze the shape of a point cloud data without actually calculating the precise geometry. It illuminates some qualitative features of  data  which persist across multiple scales. These persistent features provide an effective quantification for the shape of  data. The method is  based on techniques from algebraic topology, a branch of mathematics that deals with different ``bridges'' between algebra and topology known as ``functors". These functors take topologically equivalent (homotopic) spaces  to algebraically equivalent (isomorphic) spaces . The functorial nature of persistent homology makes it robust to perturbations of an input point cloud; a rarely found feature  in some of the existing data analysis techniques.

A number of functors exist to deal with different classes of topological spaces (simplicial, cellular, singular, etc). The development of PH is  based on the functor  known as simplicial homology. On the topological side we consider a simplicial complex $\Delta$, and on the algebraic side we get vector spaces $H_i(\Delta)$ for $i=\{0,1,2,\dots \}$. The dimension of $H_0(\Delta)$ gives the number of connected components, $H_1(\Delta)$ gives the number of holes, $H_2(\Delta)$ gives the number of voids, and so on. In PH we construct simplicial complexes $\Delta_{\epsilon}$ from a point cloud depending on a scale parameter $\epsilon$. The homological features that remain persistent across scales, provide an effective analysis of the shape of  data. The summary of these features is either shown on a ``bar diagram" (BD) or a ``persistent diagram" (PD).

In what follows we provide a formal  overview of the aforementioned terminologies. For more details see \citep{rotman2013introduction, carlsson2009topology,otter2017roadmap,edelsbrunner2000topological}.

\subsubsection{Simplicial Homology}
\begin{definition}[Simplex]
        Let $\{p_0 , p_1,\dots,p_n\}$ be an affine independent set in $\mathbb R^n$. The n-simplex generated by this set, denoted by $[p_0,p_1,\dots,p_n]$, is the convex hull of points $p_0, p_1,\dots,p_n$. Every point $x$ of this simplex can be written uniquely in the following form
        \begin{equation}
        x=\sum_{i=0}^{n} t_ip_i, \text{where} \sum_{i=0}^n t_i =1 \text{ and each } t_i \geq 0 .
        \end{equation}

\end{definition} 
\begin{figure}[!ht]
        \centering
        \fbox{\includegraphics[width=0.8\linewidth]{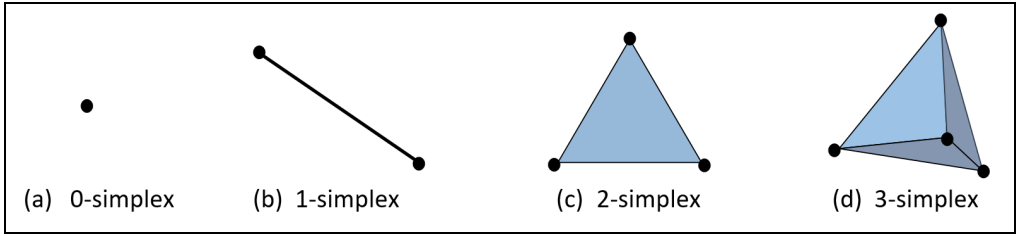}}
        \caption{Orientation of a simplicial complex}
        \label{abc}
\end{figure}

A $k$-face of $[p_0,p_1,\dots, p_n]$ is a simplex generated by a collection of $k+1$ points from $\{p_0,p_1,\dots, p_n\}$. A $k$-face is in fact a $k$-dimensional geometric object.
A simplicial complex is obtained by ``gluing" together different simplices along their common faces. 

\begin{definition}[Simplicial Complex]
        A finite simplicial complex $\Delta$ is a collection of simplices in $\mathbb R^n$ such that (1) if $\alpha \in \Delta$ then every face of $\alpha$ belong to $\Delta$, (2) for any two simplices $\alpha_1,\alpha_2 \in \Delta$, the intersection $\alpha_1\cap \alpha_2$ is either empty or a common face of $\alpha_1$ and $\alpha_2$. \end{definition} 

These conditions guarantee  that $\Delta$ records changes in each  dimension. 
In order to calculate simplicial homology we need an orientation on the simplicial complex.  An orientation of a simplicial complex is a partial order of its vertices which when restricted to a particular simplex gives a total order.

\begin{figure}[H]
        \centering
        \fbox{\includegraphics[width=0.3\linewidth]{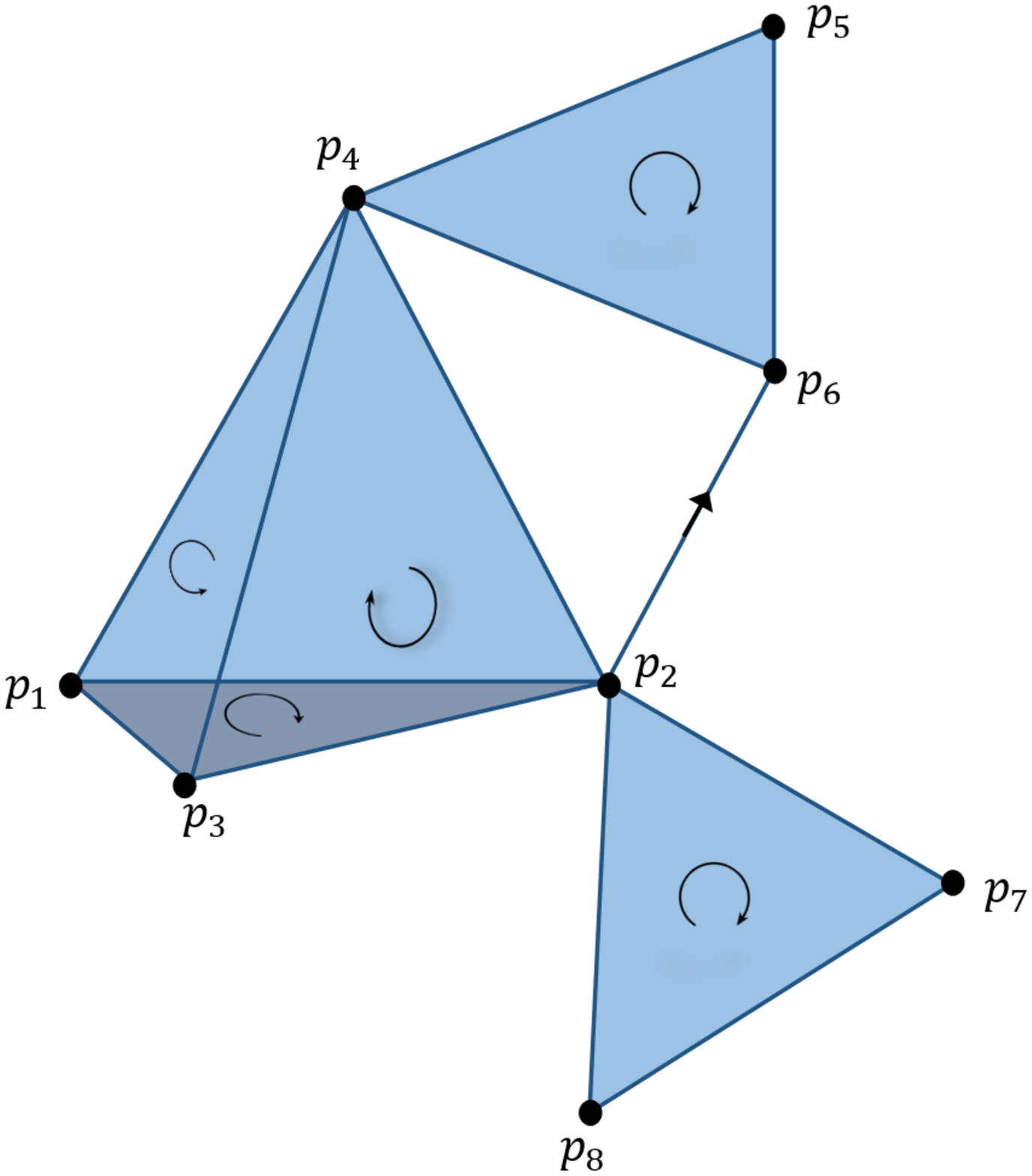}}
        \caption{A simplicial complex with a partial order}
        \caption*{$p_1<p_2<p_3<p_4<p_5<p_6$,  $p_2<p_7<p_8$ }
        \label{fig:coffee}
\end{figure}

\subsubsection{Homology of a Simplicial Complex}
For an oriented simplicial complex $\Delta$, and  integer $m\geq -1$ the $m$th chain group  $C_m(\Delta)$ consists of formal sums of the form
$$
a_1\sigma_1+a_2 \sigma_2+\dots+a_n\sigma_n, \text{ where } a_1,\dots,a_n\in \mathbb Z
$$
and $\sigma_1,\dots,\sigma_n$ are  oriented $m$-simplices of $\Delta$. For convenience, we define $C_m(\Delta)=\{0\}$ for $m> \dim \Delta$  and $m<0$. An oriented simplicial complex is an element of the chain group $C_m(K)$ of the form $\pm <p_0,p_1,\dots,p_m>$ where $p_0,\dots,p_{m}$ are distinct. The boundary operator $\partial_m: C_m(\Delta)\to C_{m-1}(\Delta)$ is defined by setting

$$
\partial_m (<p_0,p_1,\dots,p_m>)= \sum_{i=0}^m (-1)^i <p_0,\dots,\hat p_i,\dots,p_m>,
$$
where $\hat p_i$ means deleting $p_i$, and extends by linearity. Combining all information, we get the following chain complex 
$$
\cdots \xrightarrow{\partial_{m+1}} C_m(\Delta) \xrightarrow{\partial_{m}} C_{m-1}(\Delta)\xrightarrow{\partial_{m-1}}\cdots  \xrightarrow{\partial_{2}} C_1(\Delta)\xrightarrow{\partial_{1}} C_0(\Delta) \xrightarrow{\partial_{0}} 0,
$$
such that the composition of any two consecutive maps is a zero map. We define the $m$-th  simplicial homology as 
$$
H_m(\Delta) = \kernel \partial_{m} / \image (\partial_{m+1}),  \text{ where }0\leq m \leq \dim \Delta .
$$
Its dimension
$$
\beta_m(\Delta) = \dim H_m(\Delta)
$$
is called the $m$th Betti number of the simplicial complex $\Delta$. 
\subsubsection{Persistence Diagram (PD)} \label{PD}
\begin{figure}[H]
        \centering
        \includegraphics[width=16cm]{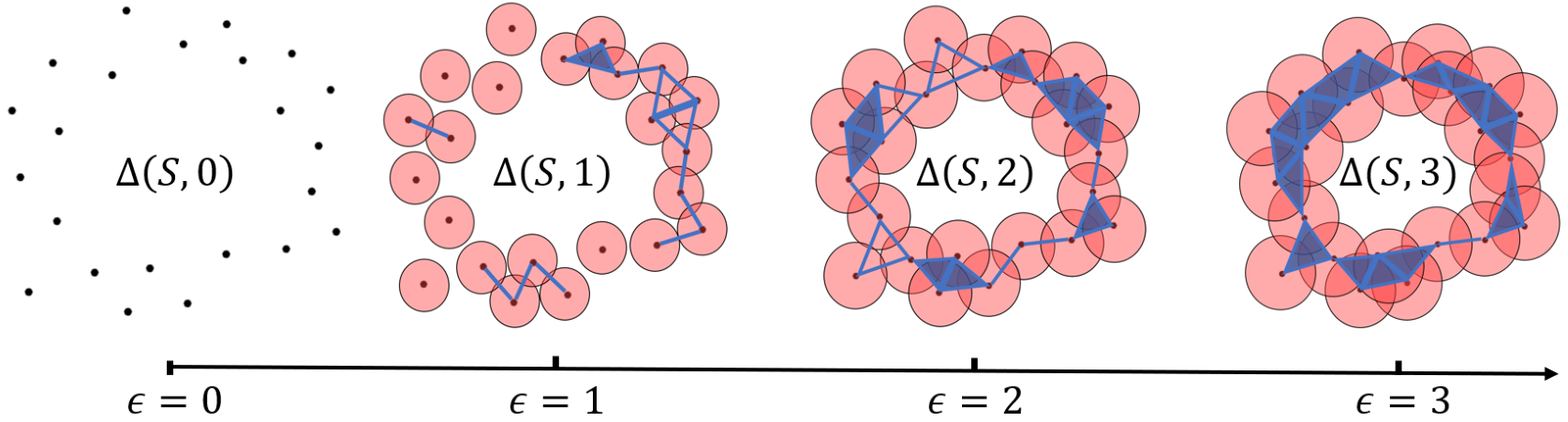}
        \caption{A filtration of simplicial complexes build from a given point cloud $\Delta(S,0)$}
        \label{CechComplex}
\end{figure}

For a data set $\mathbb X$, we can calculate its Betti numbers after imposing a simplicial complex $\Delta(X)$ on it. This information is not useful since it only gives number of connected components, 1-dimensional holes, etc. To extract further information, we construct a filtered complex, which consists of nested subcomplexes $\Delta(X,\epsilon)$  of $\Delta$, that depend on a scale parameter $\epsilon$, such that  
$$
\Delta(X,\epsilon_1)\subseteq \Delta(X,\epsilon_2) \text{ whenever } \epsilon_1 \leq \epsilon_2 .
$$
We can apply the simplicial homology functor on each subcomplex. The inclusion map on the complexes  $\Delta(X,\epsilon_1) \subset \Delta(X,\epsilon_2)$ induces a linear map on the homology groups $H_m(\Delta(X,\epsilon_1)) \to H_m(\Delta(X,\epsilon_2))$, for $0\leq m\leq \dim \Delta$. Hence the homology of this filtration complex consistently provide information about $\Delta$ at different values of $\epsilon$. 
One can represent these features, over various scales, using persistent diagrams (PDs). To exemplify, consider a point cloud in $\mathbb R^2$, say $S$, given in Fig \ref{CechComplex}. To extract qualitative information  from this data we compute  topological features for  different values of a scale parameter $\epsilon$. At each scale level of $\epsilon$, we  consider open discs, say $D(p,\epsilon)$, of radius $\epsilon$ around each point $p$. Then we build a simplicial complex $\Delta(S,\epsilon)$ using the following rule; a set of points $A=\{p_0,\dots,p_n\}$ forms an n-simplex if $\cap_{p\in A} D(p,\epsilon)\neq \Phi$. At each level the simplicial complex is made up of  simplices  shown in the Fig \ref{CechComplex}.

As the values of $\epsilon$ increase from 0, we get a filtration of simplicial complexes $\Delta(S,\epsilon)$

\begin{figure}[H]
        \centering
        \includegraphics[width=10cm]{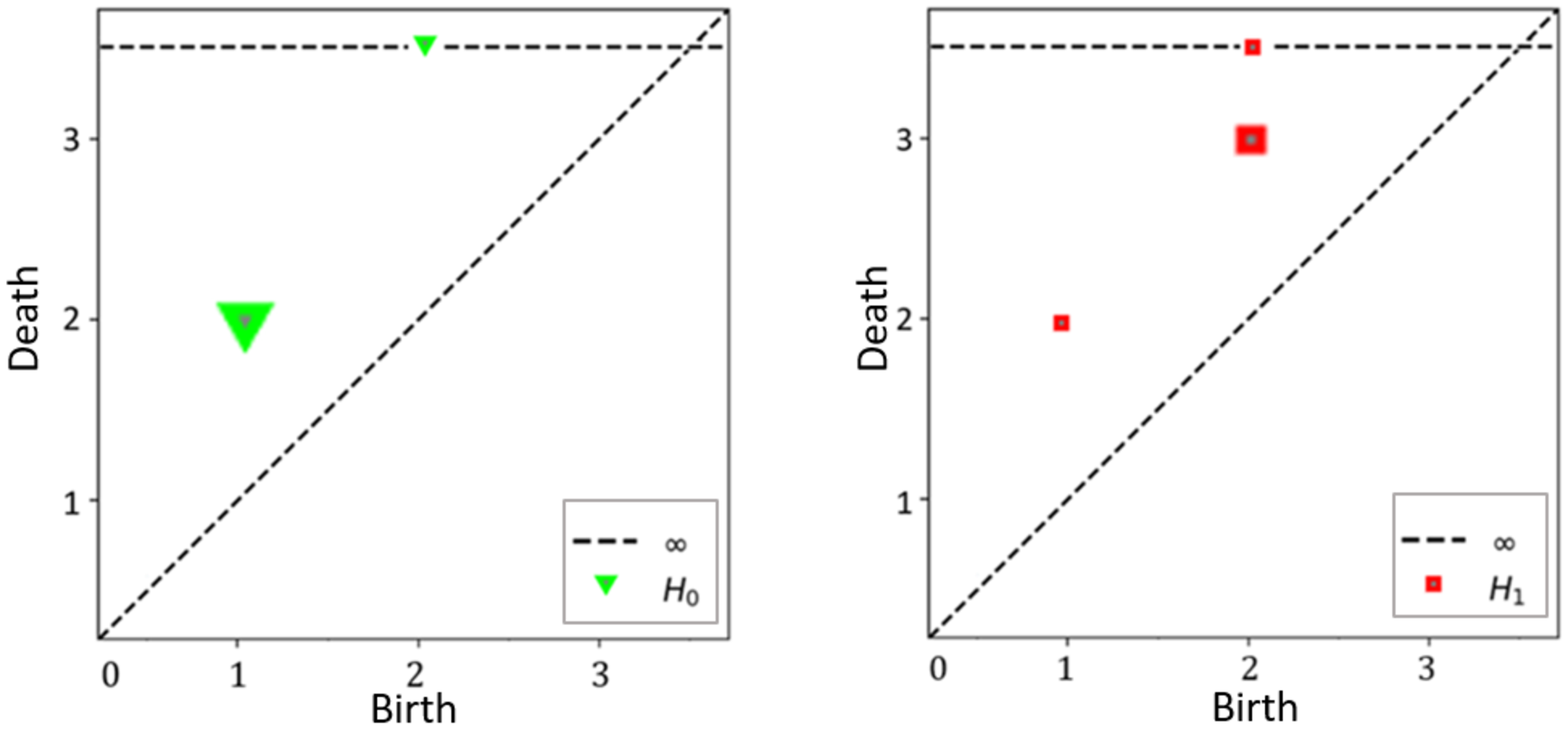}
        \caption{PDs representing topological features of the point cloud from Fig \ref{CechComplex}}
        \label{PDsofCechComplex}
\end{figure}

We can represent the birth and death of these topological features using the persistent diagram. A persistent diagram (PD) is a collection of ordered pairs in the extended plane. A point $(a,b)$ represent the birth at scale parameter value $\epsilon=a$ and death at $\epsilon =b$. The points that touches the infinity line are the persistent features that do not die till the last value of our filtration parameter. In Fig \ref{PDsofCechComplex} we see a point  at line of infinity. This depicts that there is one-dimensional hole (or loop) that appears at $\epsilon=2$ (see Fig \ref{CechComplex}) and never dies. It is very consisting with the observation that the overall shape of data in Fig \ref{CechComplex} is circular.

Apart from its stability, another important feature of the PH is that it can be computed using different algorithms. There are many libraries available that implement algorithms for the computation of PH. These libraries include, \perseus, \javaplex, \dionysus, \ripser, \gudhi, etc. We use \ripser \citep{ctralie2018ripser} due to its computational efficiency \citep{otter2017roadmap}.

There are many ways to impose a simplicial complex on a given point cloud. The choice depends on the nature of data, computational cost, and restrictions of software/package used. Some typical simplicial complexes are, Vietoris--Rips complex, \v{C}ech complex, Delaunay complex, clique complex, alpha complex, strong witness complex, weak witness complex, etc.

\section{Methods and Algorithm}\label{MaA}
\begin{figure}[H]
        \centering
        \includegraphics[width=14cm]{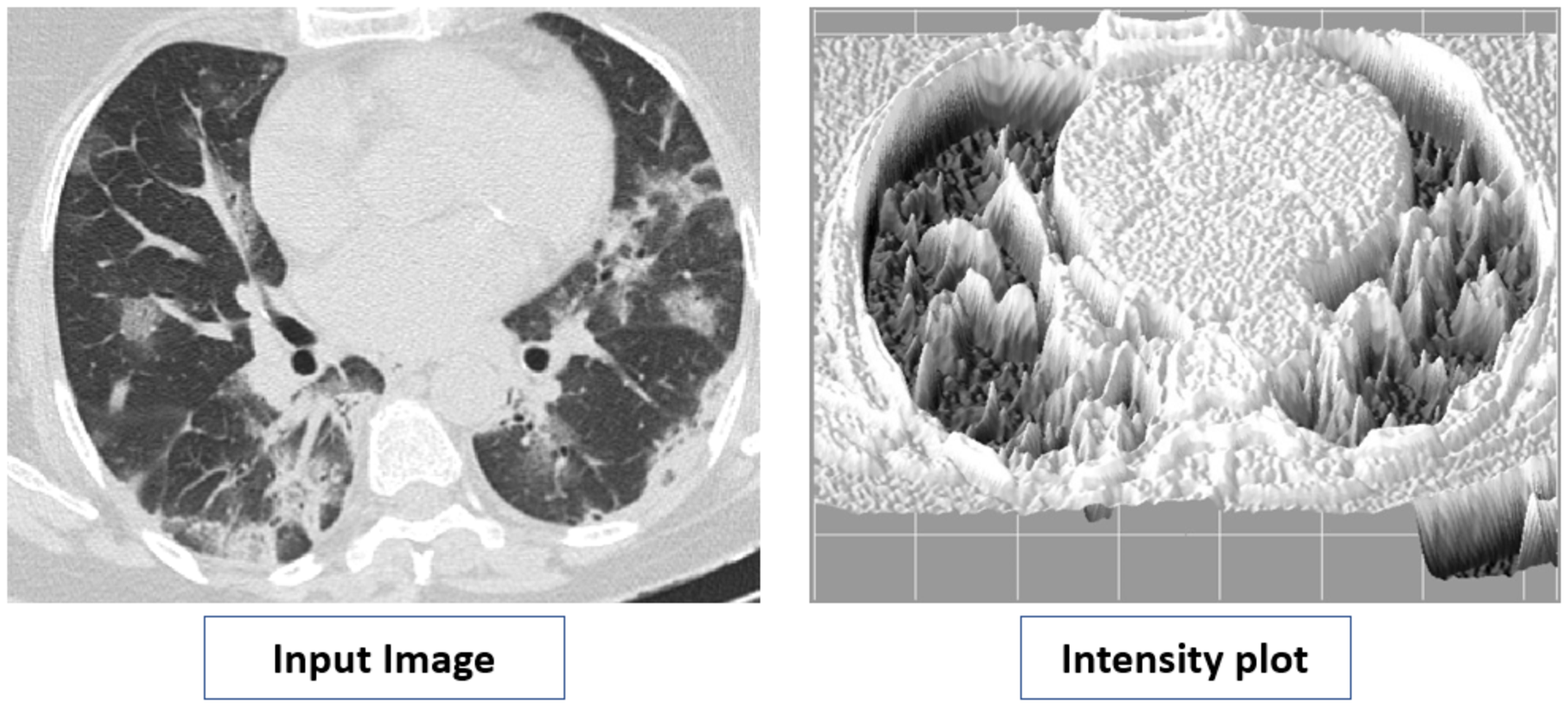}
        \caption{Shape features and intensity plot in a COVID-19 infected CT-scan}
        \label{MD}
        
\end{figure}

Thoracic radiology evaluations found high rates of ground-glass opacities and consolidations in  COVID-19 patients. One can observe the ground-glass opacities (GGOs) together with consolidations in the CT-Scan of  COVID-19 images.
These regions are isolated with difference in shapes (left image in Fig. \ref{MD}), which is captured by PDs associated to $H_0$ and $H_1$. Moreover, these regions have unique shape in the intensity plot, see appearance of alps, saddle points in Fig. \ref{MD} (right image). These are recorded by PDs of Lower-star-filtration and $H_2$.

\begin{figure}[H] 
        \centering
        \fbox{ \includegraphics[width=16cm]{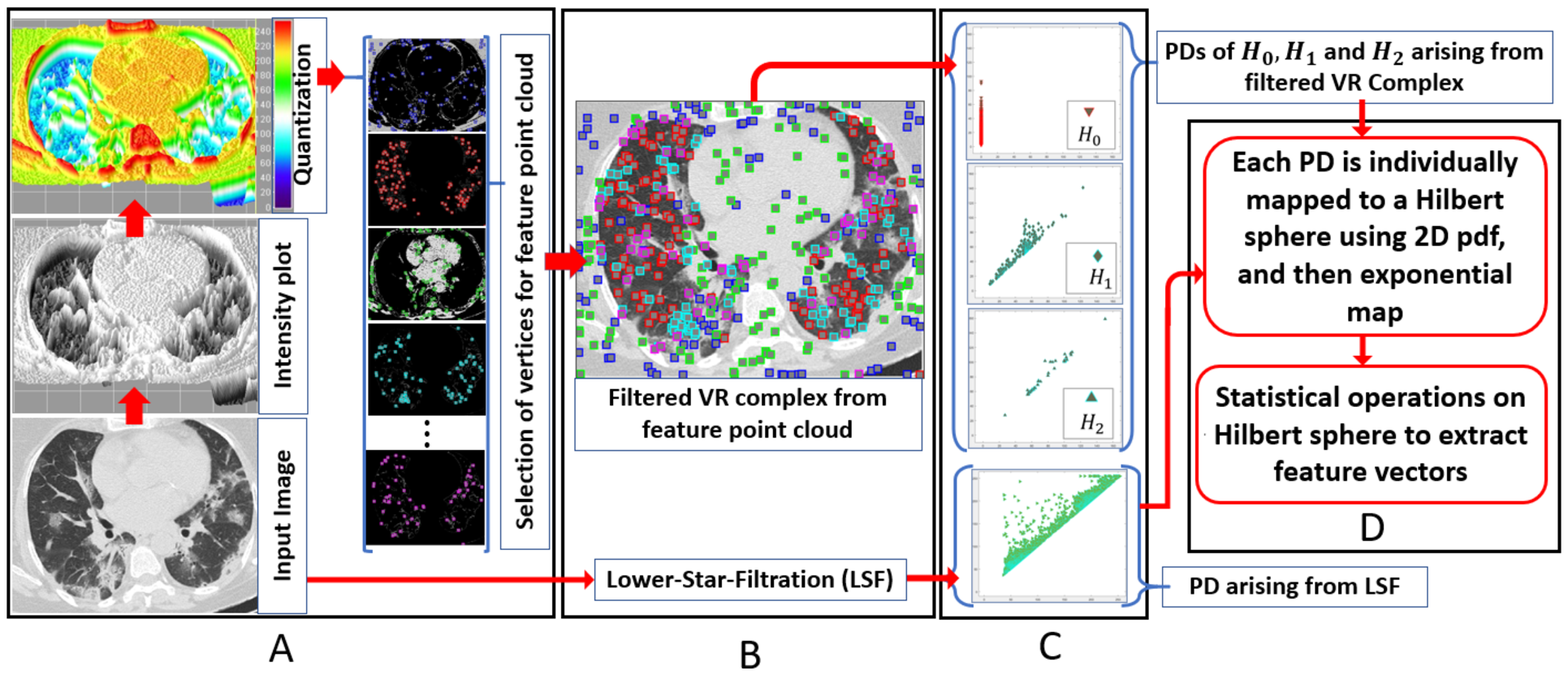}}
        \caption{Extraction of topological features from a CT-Scan of a COVID-19 patient. (A) Quantization of the Input Image which is used to construct feature point cloud (FPC). (B) Construction of two filtered complexes: filtered VR complex constructed from FPC, and lower-star filtration directly from image. (C) Calculation of PDs (D) Use of Riemannian framework to map PDs on a Hilbert sphere. Later, by using PGA, we construct a unique feature vector of the Input Image.}
        \label{MD2}
\end{figure}

All These shape features are captured by PDs associated to two different filtered complexes, namely, filtered Vietoris-Rips (VR) complex, and lower-star-filtration. The whole process is summarized as two pipelines in Fig \ref{MD2}. In the following subsections, we describe each step of these two pipelines. 

There are four PDs associated to an image; three from filtered VR complex corresponding to $H_0, H_1$ and $H_2$, and one from lower-star-filtration. 
To simplify statistical operations we map each PD onto a Hilbert sphere, this is explained in Subsection \ref{Statistics}. On Hilbert sphere,  we reduce dimension by using principal geodesic analysis (PGA)\citep{fletcher2004principal}. Eventually each PD is mapped onto a  vector of length 2400, a juxtaposition of these vectors gives a combined feature vector for each image. In the final step, in Subsection \ref{model_development}, we build a model from these feature vectors. 

\subsection{PDs Associated to a CT-Scan Image}
\subsubsection{Filtered VR Complex} \label{quantization}
Before calculating PH we first build feature point cloud (FPC) from a given point cloud. The FPC is an optimal way to record changes for all values of some chosen feature while keeping the computational time in a feasible limit. 

Let $\mathbb X$ be the point cloud with a metric $d_{\mathbb X}$, and $\mathcal F$ be a compact feature space endowed with a metric $d_{\mathcal F}$. We define the projection map $\pi : \mathbb X \to \mathcal F$ such that $\pi (p)$ is the feature of the point $p$. Let $U=\{U_i\}_{i \in I}$ be a  covering of $\mathcal F$, where $I$ is a finite indexing set, the finiteness is guaranteed by the compactness of $\mathcal F$. Using the projection map $\pi$, we get a covering $\mathcal C(\mathbb X)$ of $\mathbb X$, where $\mathcal C (\mathbb X) = \{\pi^{-1}(U_i)\}_{i\in I}$. Let $\Gamma(\mathbb X, U)$ be the connected components of $\mathcal C ( \mathbb X)$, that is,  
$$
\bar {\mathbb X}(U) =\{V:  \exists k\in I \text{ such that } V \text{ is connected
        component  of } \pi^{-1}(U_k)\}.
$$
We build a feature point cloud (FPC) by taking vertices to be points of $\bar{\mathbb X}(U)$. The distance between two connected components is taken as the distance between their centroids.

For a gray image $M$, let $\mathbb X_M$ be the point cloud in $\mathbb R^3$ defined as
$$
\mathbb X_M=\{(i,j,p): \text{ p is the intensity value at } (i,j)\}
$$
For feature space $\mathcal F_M=[0,255]$,  such that  $\mathcal \pi :\mathbb X_M \to [0,255]$ is defined as  $\mathcal \pi (i,j,p)=p$. For some finite cover $W=\{W_i\}_{i\in J}$ of $\mathcal F_M$ we get an FPC, denoted as, $\bar{\mathbb X}_M(W)$. The cover is chosen  in such a way that the resulting FPC provides a good approximation of the PH of $\mathbb X_M$. To calculate PH we use filtered Vietoris--Rips complex 
$$
VR_{0} (\bar{\mathbb X}_M(W))\subset VR_{0.5} (\bar{\mathbb X}_M(W)) \subset \dots\subset VR_{500} (\bar{\mathbb X}_M(W)).
$$
where 
$$
VR_{\epsilon} (\bar{\mathbb X}_M(W))= \left\{   A\subset \bar{\mathbb X}_M(W): d_{\mathbb X}(a_1,a_2)<2\epsilon , \, \forall a_1,a_1\in A  \right\}.
$$

\subsubsection{PDs: Capturing the Visible}
\begin{figure}[H] 
        \centering
        \fbox{ \includegraphics[width=16cm]{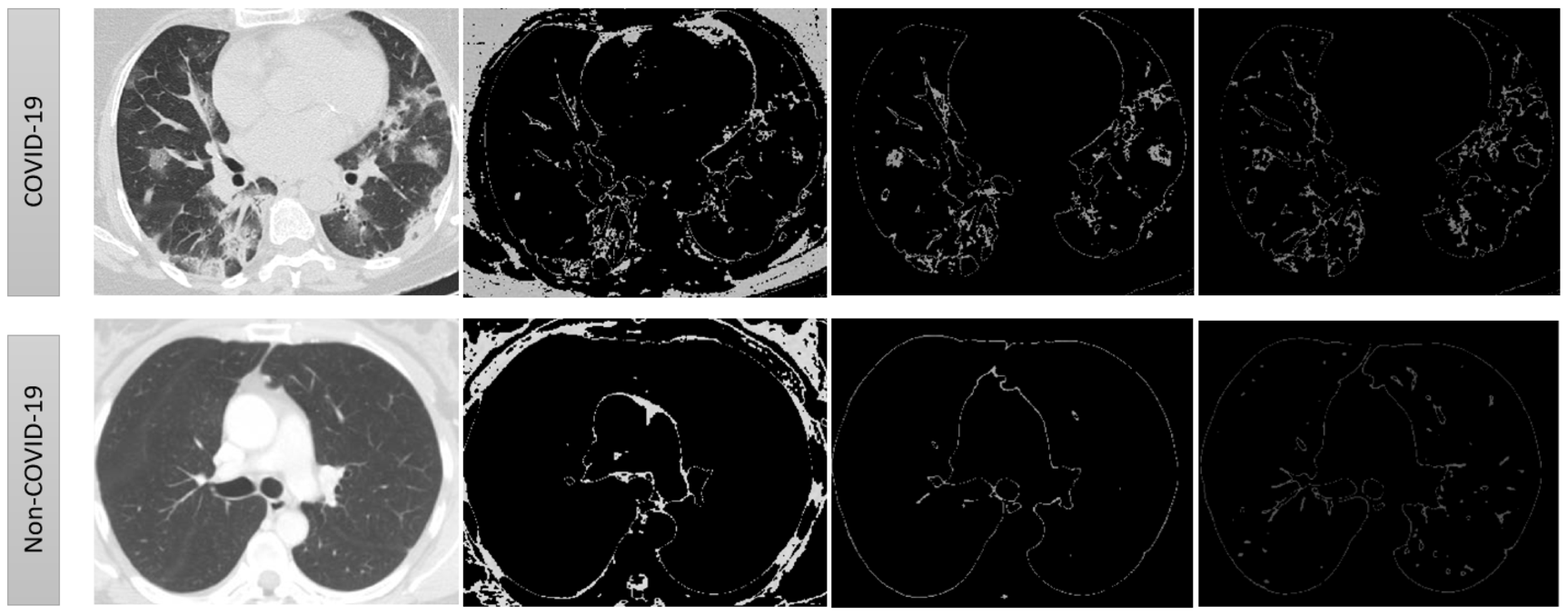}}
        \caption{Comparison of COVID and Non-COVID CT-scans. Difference in the variation of connected components ($H_0$) and (randomly shaped) loops ($H_1$) at different intensity levels can be seen.}
        \label{Loops_and_con_comp}
\end{figure}

\begin{figure}[H] 
        \centering
        \fbox{ \includegraphics[width=16cm]{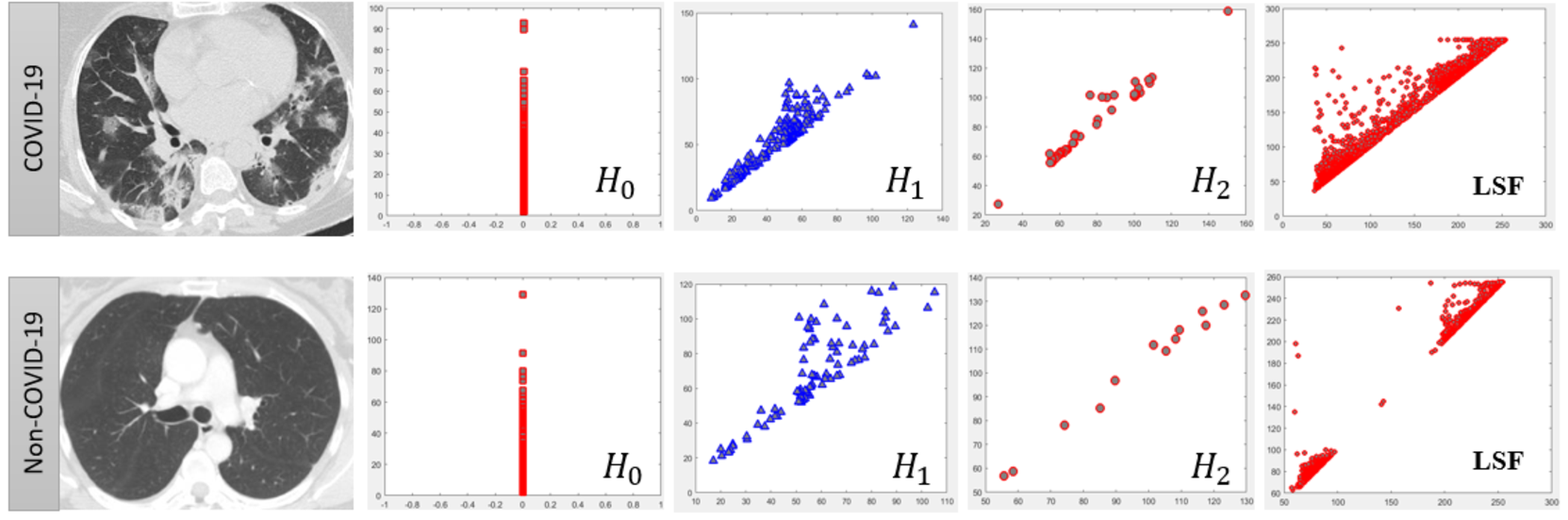}}
        \caption{PDs of $H_0, H_1,H_2$, and lower-star-filtration of a COVID-19 infected, and a Non-COVID-19 CT-scan}
        \label{PDs_of_CaNC_image}
\end{figure}

This filtered complex, defined in the previous section, is able to capture the key features of a CT-scan, like, peaks, variations in intensity, etc. Connected components and appearance of loops at different intensity levels are captured in the PD's of  $H_0$, and $H_1$ respectively (see Fig. \ref{Loops_and_con_comp}). The PDs of $H_0$ and $H_1$ for the images in Fig. \ref{Loops_and_con_comp} are given in Fig. \ref{PDs_of_CaNC_image}. It is evident that the difference in their visual appearance is captured by these PDs.  

\begin{figure}[H] 
        \centering
        \fbox{ \includegraphics[width=16cm]{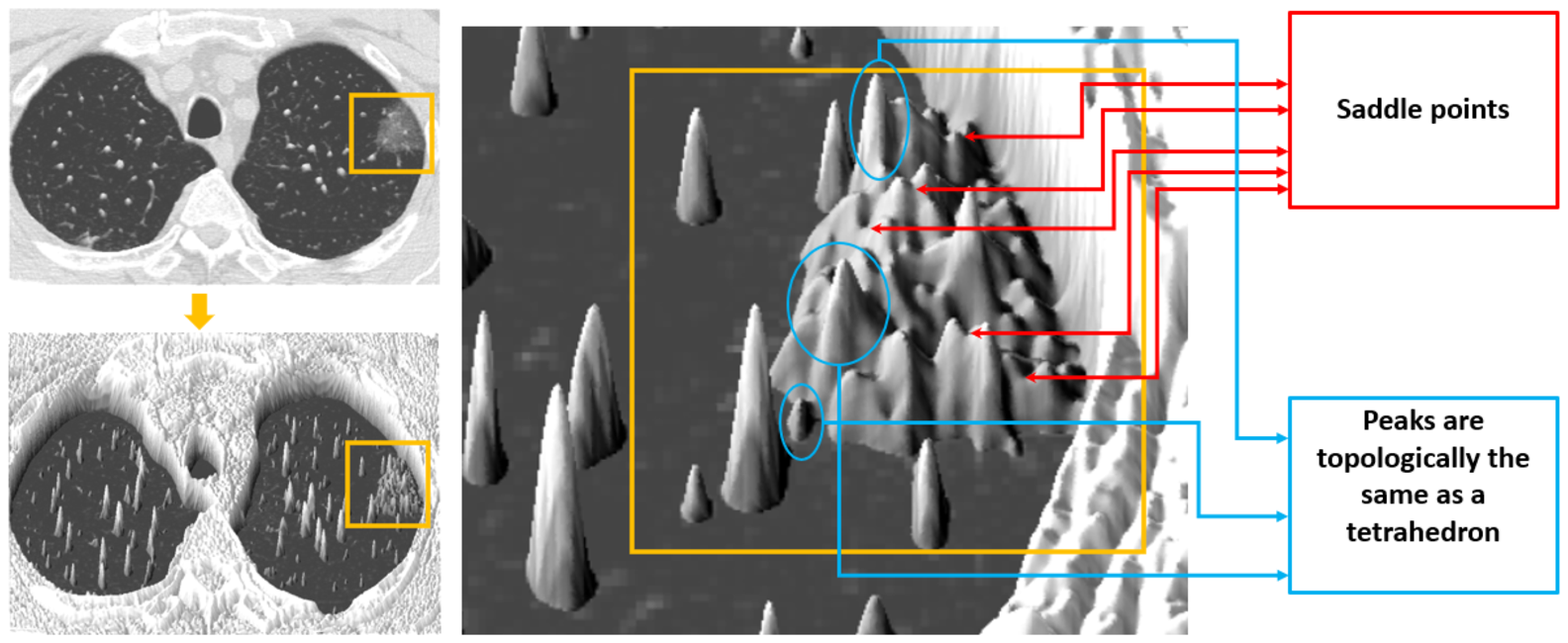}}
        \caption{Peaks are captured by VR complex and saddle points are captured by lower-star-filtration}
        \label{peaks_and_saddle}
\end{figure}

A peak, which is topologically equivalent to tetrahedron, appear as points at (or near to) infinity-line in the PD of $H_2$, that is, a persistent 2-dimensional void (see Fig. \ref{peaks_and_saddle}, \ref{PDs_of_CaNC_image}).

 Lower Star Filtration (LSF) captures key features about variation in intensities in an image. Local minimum and saddle points, in the intensity plot, are vital shape features (see Fig \ref{peaks_and_saddle}). These features are recorded using LSF. The birth time of a point in this PD   is local minimum and death time is saddle point.

To record these changes, we construct a simplicial complex, say $\Delta$, in the following way. Each pixel is taken as a vertex, and there is an edge from one vertex to its neighbouring 8 (or less in case its an edge vertex) vertices. This allow us to construct the  filtration
$
\Delta_{0}\subset \Delta_{1} \subset \dots \subset \Delta_{255},
$
where $\Delta_{\epsilon}=\{ A\subset \Delta : \max_{a\in A} p(a) \leq \epsilon \},$ here $p(a)$ is the pixel value of the vertex $a$. Only the zero-dimensional PD is essential for our model development. The Fig.\ref{PDs_of_CaNC_image} shows a major difference between the PDs associated to LSF of a COVID-19 and a non-COVID-19 CT-scan image. 

\subsection{Statistics on Space of Persistent Diagrams} \label{Statistics}
\begin{figure}[H] 
        \centering
        \fbox{ \includegraphics[width=12cm]{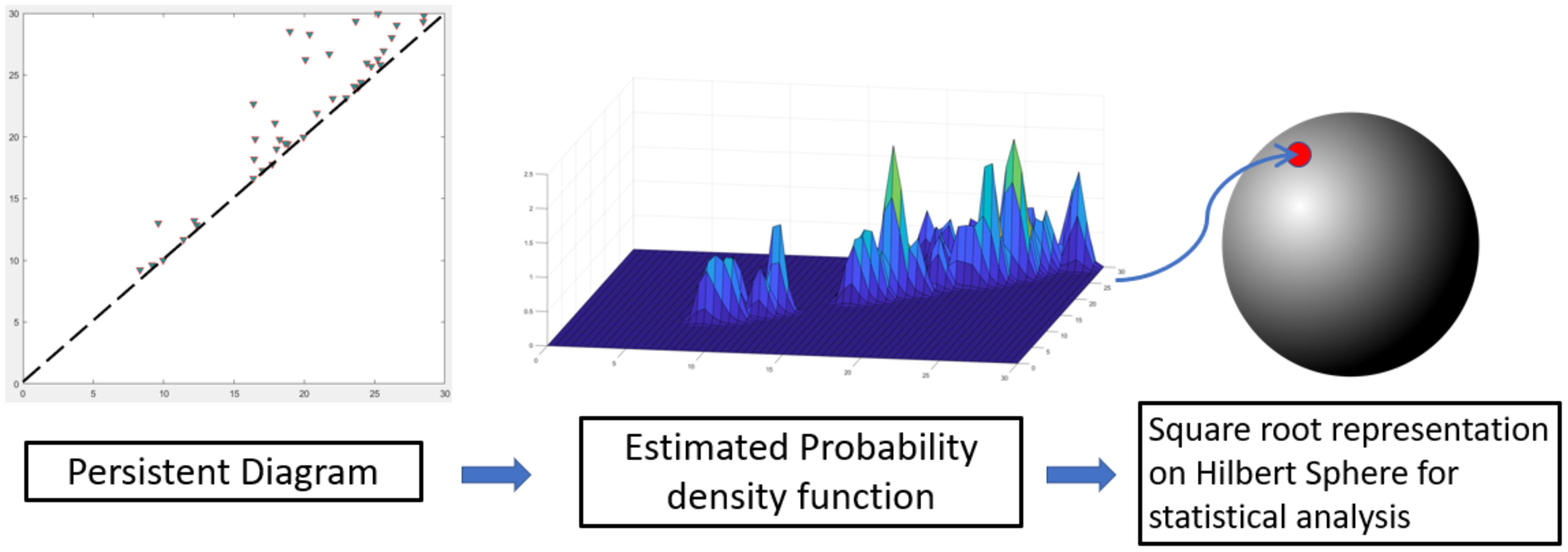}}
        \caption{Riemannian framework approach estimates a PD with a pdf, and then maps onto a Hilbert sphere}
        \label{RD}
\end{figure}

There are many approaches to infer results from a collection of PD's, for example, bottleneck distance \citep{cohen2007stability} (and its generalizations), $L_{p}$-Wasserstein metric \citep{cohen2010lipschitz}, persistent landscape \citep{bubenik2015statistical}, and Riemannian framework \citep{anirudh2016riemannian}, etc. Due to its efficiency and computational cost, we use Riemannian framework to perform our statistical analysis, which includes principal geodesic analysis, and SVM  on the space of PD's. In this approach (see figure \ref{RD}) we first approximate a given PD with a 2D probability distribution function (pdf) that are further mapped, using square-root transformation, onto a Hilbert sphere. On Hilbert sphere we have closed-form expressions to compare two PD's. Using  this we first apply principal geodesic analysis to reduce our dimensions., and then we build our model using SVM.

For each point $(a,b)$ of a PD, we use multivariate normal distribution with parameters $\Sigma=\begin{pmatrix}0.2 & 0 \\
0 & 0.2 \\
\end{pmatrix}$ and $\mu= (a,b)$. We calculate the values of this 2D pdf on the meshgrid $[0,530]^2$ with a uniform difference   of $0.5$. Further, we compute square-root representation of this pdf which maps them on a Hilbert sphere (c.f. \cite[Section 3.3]{anirudh2016riemannian}). Hence each PD is converted into a vector of length $1061^2$. To reduce this dimension we apply Principal Geodesic Analysis (PGA) \citep{fletcher2004principal} on our Hilbert sphere.  This whole procedure is applied on each collection of PD's for  $H_0, H_1,H_2,$ and on the PD coming from lower-star-filtration. Hence for a fixed image, each of the four PDs is represented as a vector of length $2400$. 

\subsection{Model Development} \label{model_development}
We build the classification  model using SVM which tends to perform relatively well on limited training data sets with high dimensional features. Each of the $2,480$ CT-Scans is represented by a homology feature vector. We use Principal Component Analysis  to reduce the dimension of  each feature vector to 4800. Before training of the SVM, we randomly split the data in to $k$-folds (where $k=5$);  we used $(k-1)$-folds for training of the SVM and the remaining $1$-fold for testing stage to validate the performance of the model on a completely unseen data set. The proposed model, using topological features of $H_0, H_1, H_2$ and LSF, achieves the classification accuracy of $99.4164 \%$ on the test data set.

We use the Radial Basis Function (RBF) kernel for the SVM model and we further select the optimal hyperparameters using the Bayesian optimization. In particular, we perform the optimization search to minimize the cross-validation loss (error) by varying the kernel scale and  box constraint. The kernel scaling parameter applies on the input features before computing the Gram matrix and the box constraint acts as a regularizer to mitigate the impact of overfitting by penalizing the margin-violating observations. For both parameters, the model searches for log-scaled positives ranging between $1e^{-3}$ and $1e^{3}$.

\section{Results and Discussion} \label{RaD}
We use a publicly available data set ``SARS-CoV-2 CT-scan dataset", which contains 1252 CT scans that are positive, and 1230 CT scans for patients non-infected for COVID-19 infection. These data is collected from patients in hospitals from Sao Paulo, Brazil and made public in \citep{angelov2020explainable,soares2020sars}.

We propose a model based on topological features, these provides a binary classification for COVID and Non-COVID CT-Scans. Our model is an attempt to capture the features as observed by a professional medic. These features are picked up by the topological summaries provided by PDs.  Hence making it  biologically more interpretable as compared to deep neural networks. Moreover topological techniques do not need plenty of data to train a model. Table \ref{Thecomparison} compare the average values of the evaluation metrics achieved by different deep networks and our topological model.

\begin{table}[H]  
        \begin{adjustbox}{width=475pt,center}  
                
                \begin{tabular}{|l||*{5}{c|}}\hline 
                        \backslashbox{Method}{Metric}
                        &\makebox[5em]{Accuracy}&\makebox[5em]{Precision}&\makebox[5em]{Recall}
                        & \makebox[5em]{Specificity} &\makebox[5em]{F1 Score} \\\hline\hline
                        Topological Approach & $99.416\pm 0.27$  & $99.41\pm 0.24$ & $99.416\pm 0.62$ & $99.424\pm 0.22$ & $99.42\pm 0.26$  \\\hline
                        
                        ResNet101 \citep{alshazly2020explainable}  &  $99.4\pm 0.4$ & $99.6\pm 0.3$& $99.1\pm 0.6$ &$99.6\pm0.3$  & $99.4\pm 0.4$  \\\hline
                        
                        SqueezeNet \citep{alshazly2020explainable} &  $95.1\pm 1.3$ & $94.2\pm 2.0$ & $96.2\pm 1.4$ & $94.0\pm2.2$ & $95.2\pm 1.2$ \\\hline
                        
                        ResNeXt101 \citep{alshazly2020explainable} &  $99.2\pm 0.3$ & $99.2\pm 0.4$& $99.3\pm 0.5$ & $99.2\pm0.4$ & $99.2\pm 0.3$  \\\hline

                        ShuffleNet \citep{alshazly2020explainable} &  $97.5\pm 0.8$ & $96.1\pm 1.4$& $99.0\pm 0.2$ & $95.9\pm1.5$ & $97.5\pm 0.8$  \\\hline

                        InceptionV3  \citep{alshazly2020explainable} &  $99.1\pm 0.5$ & $98.5\pm 0.8$ & $99.8\pm 0.3$ & $98.5\pm0.8$ & $99.1\pm 0.5$  \\\hline
                        
                        ResNeXt50 \citep{alshazly2020explainable} &  $99.1\pm0.5$ & $99.0\pm0.5$ & $99.3\pm0.5$ & $98.9\pm0.6$ & $99.1\pm0.5$ \\\hline
                        
                        DenseNet169 \citep{alshazly2020explainable} &  $99.3\pm0.5$ & $99.4\pm0.6$ & $99.3\pm0.5$ & $99.3\pm0.7$ & $99.3\pm0.4$ \\\hline
                        
                        DenseNet201 \citep{alshazly2020explainable}& $99.2\pm0.2$ & $99.4 \pm0.4$  &$99.4\pm0.2$  &$98.9\pm0.4 $& $99.2\pm0.2$  \\\hline

                        xDNN  \citep{soares2020sars} & 97.38 & 99.16 &95.53 &  -& 97.31   \\\hline
                        
                        Contrastive Learning \citep{wang2020contrastive} & $90.8\pm 0.9$ & $95.7\pm0.4$  & $85.8\pm1.1$ & - & $90.8\pm1.3$  \\\hline

                        Modified VGG19 \citep{panwar2020deep} & 95.0 &  95.3 & 94.0  &  94.7 &  94.3  \\\hline
                        
                        COVID CT-Net \citep{yazdani2020covid} & - & -  & $85.0\pm 0.2$  & $96.2\pm0.1$  &  $90.0 \pm0.1$  \\\hline
                        
                        DenseNet201 \citep{jaiswal2020classification} & 96.2 & 96.2  &  96.2 & 96.2  & 96.2   \\\hline
                \end{tabular}
                
        \end{adjustbox}

        \caption{The topological model performs better in terms of all metrics than the other state-of-the-art approaches including the latest deep learning models. See \citep{alshazly2020explainable} for the definitions of evaluation metrics.}
        \label{Thecomparison}
\end{table}

\begin{figure}[H] 
        \centering 
        \fbox{\includegraphics[width=15cm]{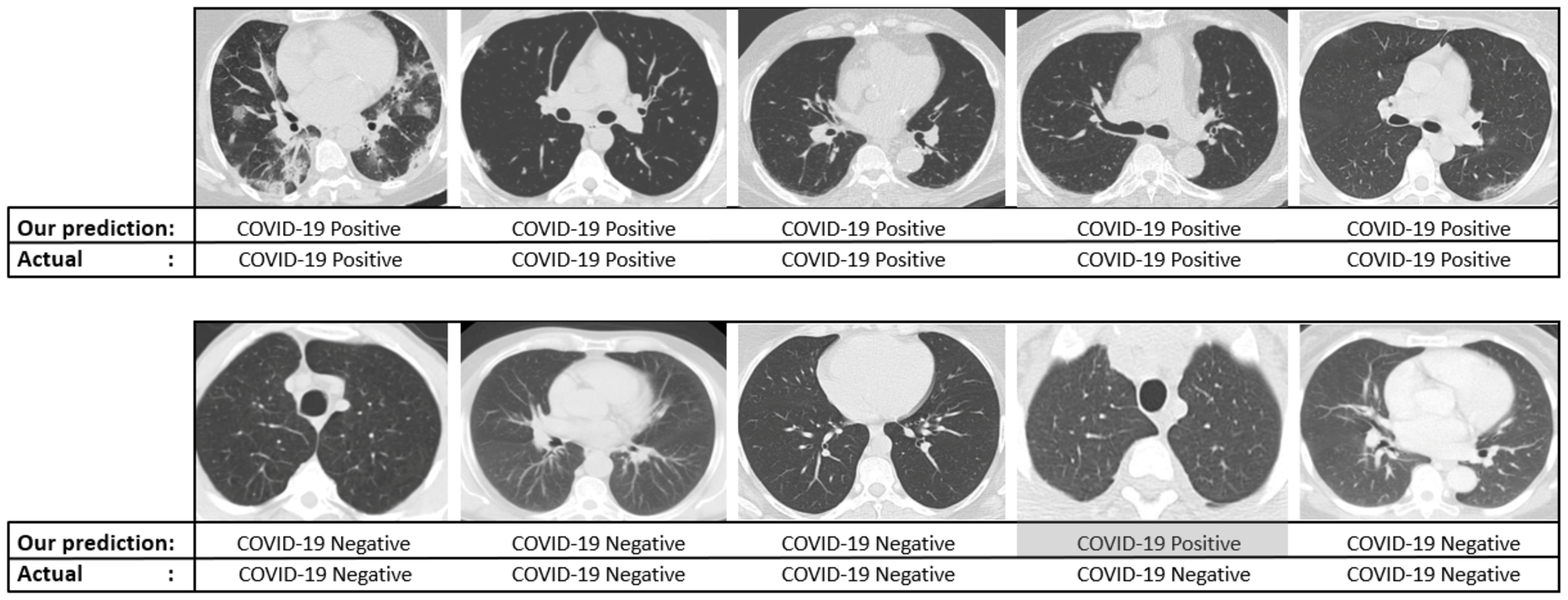}}
        \caption{Some samples of COVID-19 and Non-COVID-19 CT-Scans with the predictions of our topological model.}
        \label{Predictions}
\end{figure}

Some of the topological features are more important than the others, depending on the input image. The following table present performance of our model by using individual  topological features.

\begin{table}[h]
        \begin{tabular}{|l||*{5}{c|}}\hline
                \backslashbox{Top. Feature}{Metric}
                &\makebox[5em]{Accuracy}&\makebox[5em]{Precision}&\makebox[5em]{Recall}&\makebox[5em]{Specificity}
                &\makebox[5em]{F1 Score}\\\hline\hline
                $H_0$         &  $71.6 \pm 2.1$  &  $70.3 \pm 3.3$ &   $74 \pm 1.1$  & $69.2 \pm 3.7$ & $72 \pm 1.93$  \\\hline
                $H_1$         & $68.5 \pm 0.5 $ & $67.1 \pm 2.6$  & $72.6 \pm 7.7$  & $64.4 \pm 8.1$ & $ 69.5 \pm 32.3$  \\\hline
                $H_2$         &  $64.8 \pm 2.1$  &  $63.5 \pm 1.8 $& $67.7 \pm4.7$  & $61.9 \pm 2.2$ &  $65.5 \pm 2.5 $\\\hline
                LSF           &  $97.5\pm 0.6$  &  $97\pm 1$    & $98\pm 0.6$   & $97\pm 1$  & $97.5\pm0.6 $\\\hline
        \end{tabular}
        \caption{Performance metrics for different combinations of topological features}
        \label{TF_comparison_1}
\end{table}

Although LSF feature performs exceptionally well as compared to other features but in some situations lack of saddle points can make it less reliable. So combining all the features makes our model more reliable in all situations.  

\begin{figure}[H] 
        \centering
        \fbox{ \includegraphics[width=12cm]{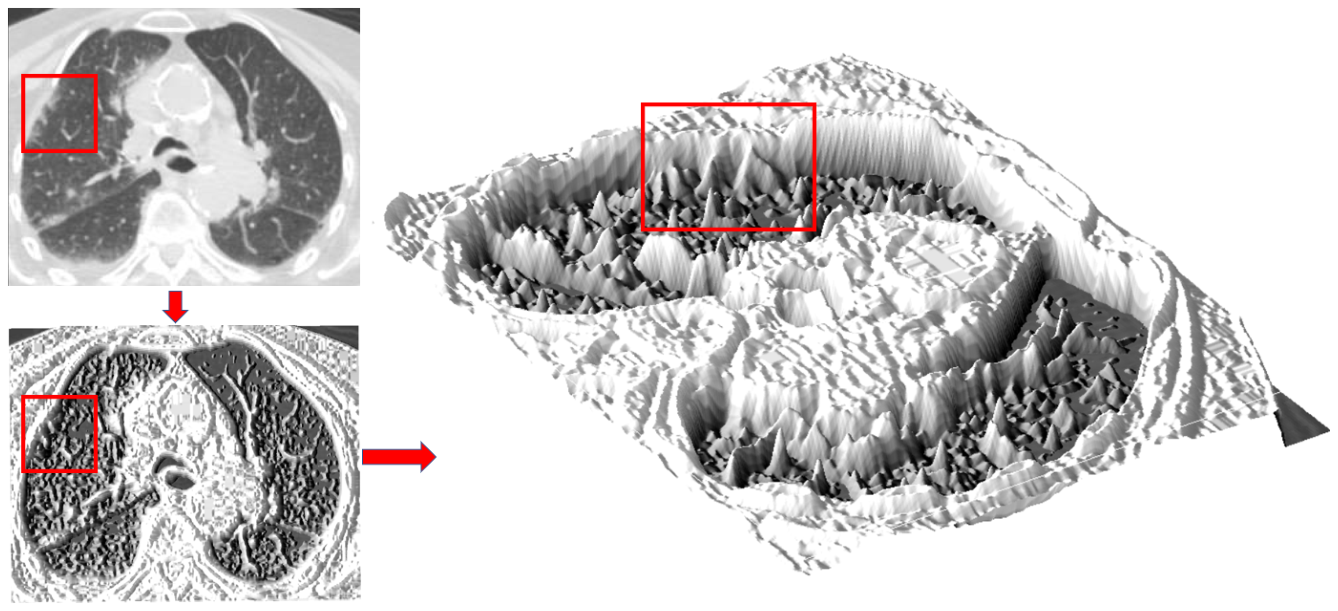}}
        \caption{A model using only LSF feature was not able to predict the traces of COVID-19 due to scarcity of saddle points in the infected region. But all models using all/one of the features $H_0,H_1, \text{ and }H_2$ detected it correctly.}
        \label{WhyNotOnlyLSF}
\end{figure}

\subsection{t-Distributed Stochastic Neighbor Embedding Visualization}
\begin{figure}[H] 
        \centering
        \fbox{ \includegraphics[width=10cm]{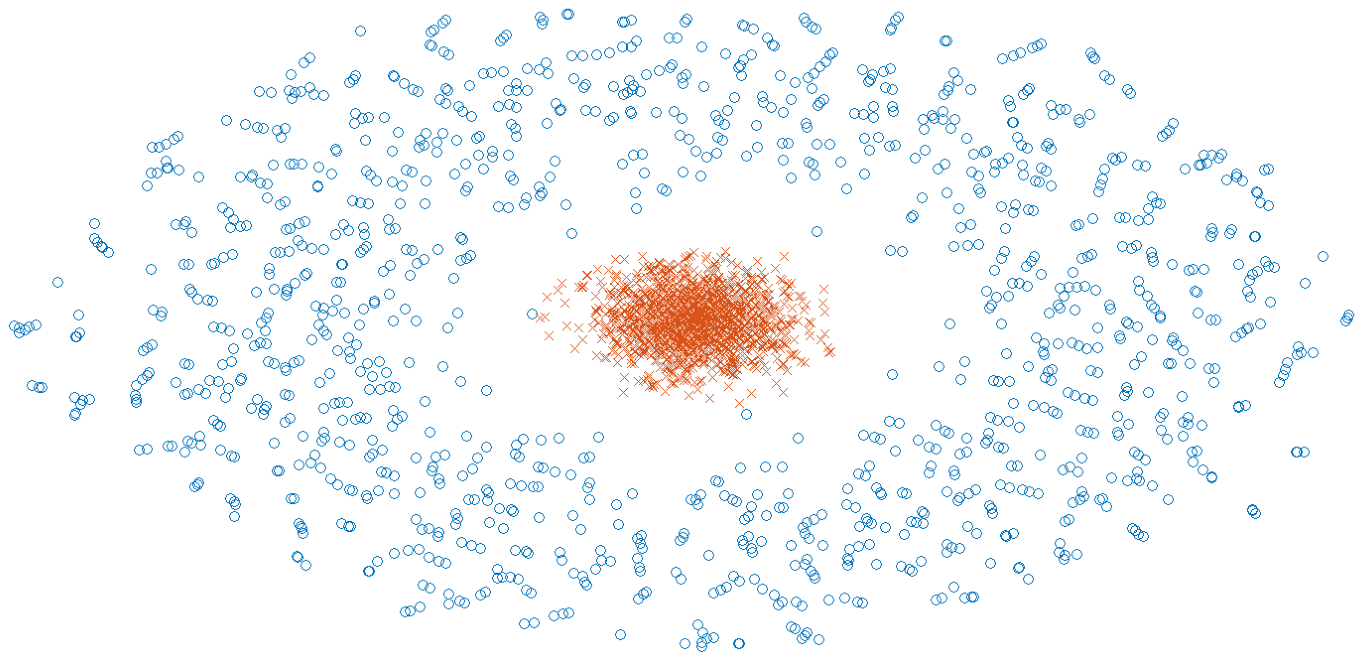}}
        \caption{Visualization of the t-SNE embedding of the entire SARS-CoV-2 CT-Scan dataset. We clearly see two different clusters representing COVID-19 (red) and Non-COVID-19 (blue). It is clear that the embedding has some of the geometric properties of Hilbert sphere.}
        \label{tsne}
\end{figure}

To visualize the relative position of topological feature vectors coming from the CT-scans of SARS-CoV-2 we apply t-SNE. In this process we take vectors of length 4800 and map them to 2D. In Fig \ref{tsne} we can clearly see two segregated clusters of the COVID-19 and Non-COVID-19 images. 

\subsection{Robustness Analysis}

Machine learning algorithms have achieved persuasive
performance in several medical imaging problem but the interpretability of these ML models is very limited and it remains a significant hurdle in adoption of these models in clinical practice. We perform  an experiment to check the robustness of our model we perform the following procedures. 

\begin{figure} [H]
        \centering
        \fbox{ \includegraphics[width=14cm]{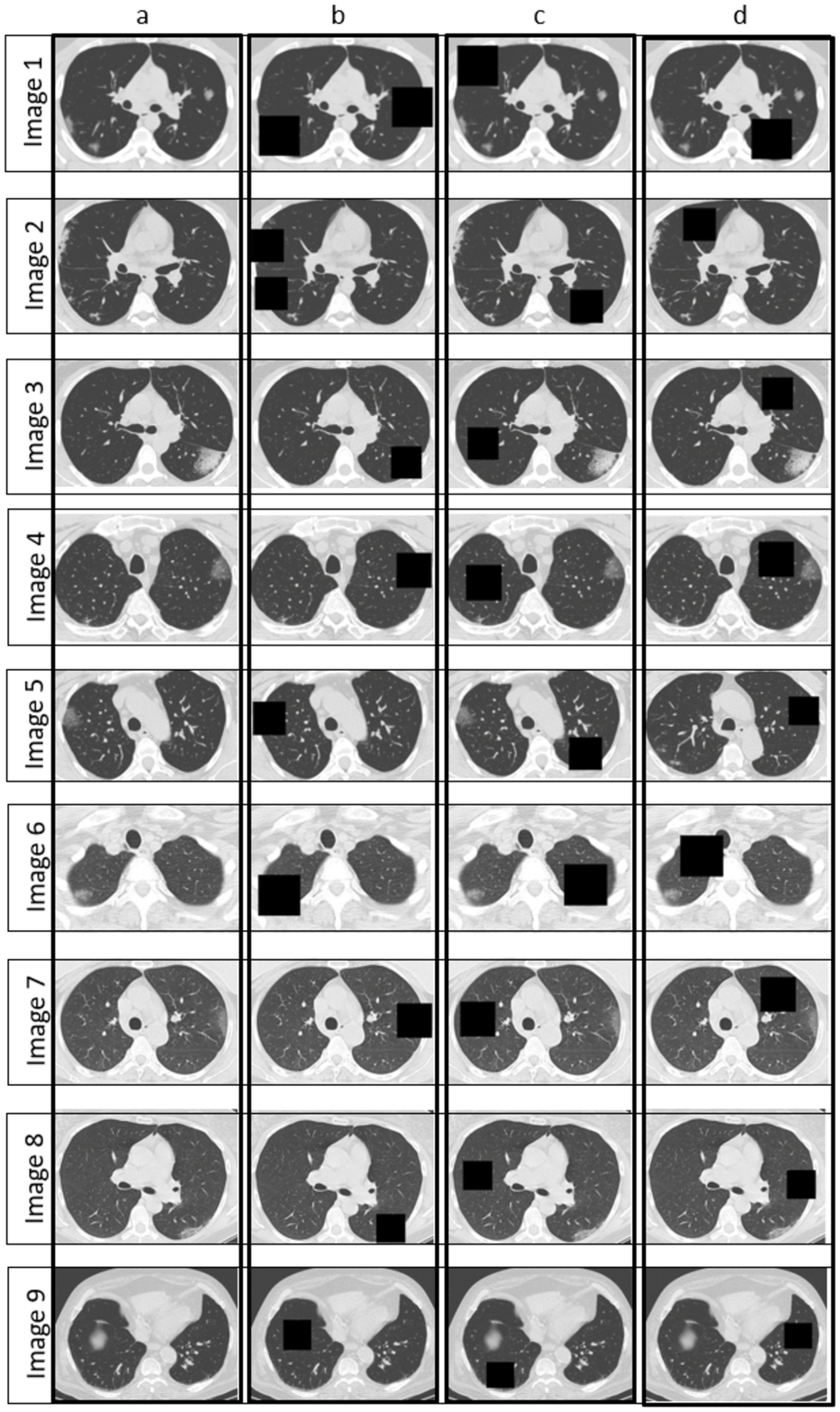}}
        \caption{(A) Original images. (B) We covered COVID-19 regions. (C)\&(D) We covered random non-COVID-19 regions. }
        \label{HeatMaps}
\end{figure}

In the first stage, we  removed critical regions (GGOs and consolidations) of interest from COVID-19 positive cases and then predict its outcome. Secondly, we investigated the performance of our model on images by randomly removing non-COVID-19 regions. We have illustrated some COVID-19 cases in Fig. \ref{HeatMaps}, where we covered the GGOs and consolidations and the model predicted it to be non-COVID-19. 

Hence the infected COVID-19 regions are very accurately captured by our chosen topological features and deductively by our model.

\section{Conclusion}\label{Con}
This paper presents a new approach to detect COVID-19 from CT-scan images, using persistent homology, and achieved state-of-the-art results. The work shows that the techniques of topological data analysis are effective and perform better that most of the deep neural networks.

This work provides a highly interpretable model based on topological features of CT-scans. The model is based on the slogan ``mimic a professional medic''. This outperforms most of cutting edge deep neural network approaches. However it will be interesting to combine topological features of this work with deep convolutional neural networks, this is left as an open direction. Chest CT-scan imaging has high sensitivity for diagnosis of COVID-19 so this is step forward in detecting and hence eliminating COVID-19.

\section*{Acknowledgments}
The first author is thankful to David Epstein and Saqlain Raza for many fruitful discussions.

\bibliographystyle{authordate1}

\bibliography{References}

\end{document}